\documentclass[]{aastex631}
\usepackage{amsmath}
\usepackage{comment}
\usepackage{textgreek}
\usepackage{multirow}

\shorttitle{}
\shortauthors{Kato et al., 2024}
\graphicspath{{./}{figures/}}

\begin{document}

\title{Quantitative constraint on the contribution of resolved gamma-ray sources to the sub-PeV Galactic diffuse gamma-ray flux measured by the Tibet AS$\gamma$ experiment}

\author{S. Kato}
\altaffiliation{Corresponding author}
\affiliation{Institute for Cosmic Ray Research, University of Tokyo, Kashiwa 277-8582, Japan; katosei@icrr.u-tokyo.ac.jp}
\author{M. Anzorena}
\affiliation{Institute for Cosmic Ray Research, University of Tokyo, Kashiwa 277-8582, Japan; katosei@icrr.u-tokyo.ac.jp}
\author{D. Chen}
\affiliation{National Astronomical Observatories, Chinese Academy of Sciences, Beijing 100012, China}
\author{K. Fujita}
\affiliation{Institute for Cosmic Ray Research, University of Tokyo, Kashiwa 277-8582, Japan; katosei@icrr.u-tokyo.ac.jp}
\author{R. Garcia}
\affiliation{Institute for Cosmic Ray Research, University of Tokyo, Kashiwa 277-8582, Japan; katosei@icrr.u-tokyo.ac.jp}
\author{J. Huang}
\affiliation{Key Laboratory of Particle Astrophysics, Institute of High Energy Physics, Chinese Academy of Sciences, Beijing 100049, China}
\author{G. Imaizumi}
\affiliation{Institute for Cosmic Ray Research, University of Tokyo, Kashiwa 277-8582, Japan; katosei@icrr.u-tokyo.ac.jp}
\author{T. Kawashima}
\affiliation{Institute for Cosmic Ray Research, University of Tokyo, Kashiwa 277-8582, Japan; katosei@icrr.u-tokyo.ac.jp}
\author{K. Kawata}
\affiliation{Institute for Cosmic Ray Research, University of Tokyo, Kashiwa 277-8582, Japan; katosei@icrr.u-tokyo.ac.jp}
\author{A. Mizuno}
\affiliation{Institute for Cosmic Ray Research, University of Tokyo, Kashiwa 277-8582, Japan; katosei@icrr.u-tokyo.ac.jp}
\author{M. Ohnishi}
\affiliation{Institute for Cosmic Ray Research, University of Tokyo, Kashiwa 277-8582, Japan; katosei@icrr.u-tokyo.ac.jp}
\author{T. Sako}
\affiliation{Institute for Cosmic Ray Research, University of Tokyo, Kashiwa 277-8582, Japan; katosei@icrr.u-tokyo.ac.jp}
\author{T. K. Sako}
\affiliation{Nagano Prefectural Institute of Technology, Ueda, 386-1211, Japan}
\author{F. Sugimoto}
\affiliation{Institute for Cosmic Ray Research, University of Tokyo, Kashiwa 277-8582, Japan; katosei@icrr.u-tokyo.ac.jp}
\author{M. Takita}
\affiliation{Institute for Cosmic Ray Research, University of Tokyo, Kashiwa 277-8582, Japan; katosei@icrr.u-tokyo.ac.jp}
\author{Y. Yokoe}
\affiliation{Institute for Cosmic Ray Research, University of Tokyo, Kashiwa 277-8582, Japan; katosei@icrr.u-tokyo.ac.jp}

\begin{abstract}
  Motivated by the difference between the fluxes of sub-PeV Galactic diffuse gamma-ray emission (GDE) measured by the Tibet AS$\gamma$ experiment and the LHAASO collaboration, our study constrains the contribution to the GDE flux measured by Tibet AS$\gamma$ from the sub-PeV gamma-ray sources in the first LHAASO catalog plus the Cygnus Cocoon. After removing the gamma-ray emission of the sources masked in the observation by Tibet AS$\gamma$, the contribution of the sources to the Tibet diffuse flux is found to be subdominant; in the sky region of $25^{\circ}<l<100^{\circ}$ and $|b|<5^{\circ}$, it is less than $26.9\%\pm 9.9\%$, $34.8\%\pm 14.0\%$, and $13.5\%^{+6.3\%}_{-7.7\%}$ at $121\, {\rm TeV}$, $220\, {\rm TeV}$, and $534\, {\rm TeV}$, respectively. In the sky region of $50^{\circ}<l<200^{\circ}$ and $|b|<5^{\circ}$, the fraction is less than $24.1\%\pm 9.5\%$, $27.1\%\pm 11.1\%$ and $13.5\%^{+6.2\%}_{-7.6\%}$. After subtracting the source contribution, the hadronic diffusive nature of the Tibet diffuse flux is the most natural interpretation, although some contributions from very faint unresolved hadronic gamma-ray sources cannot be ruled out. Different source-masking schemes adopted by Tibet AS$\gamma$ and LHAASO for their diffuse analyses result in different effective galactic latitudinal ranges of the sky regions observed by the two experiments. Our study concludes that the effect of the different source-masking schemes leads to the observed difference between the Tibet diffuse flux measured in $25^{\circ}<l<100^{\circ}$ and $|b|<5^{\circ}$ and LHAASO diffuse flux in $15^{\circ}<l<125^{\circ}$ and $|b|<5^{\circ}$.
\end{abstract}


\section{Introduction} \label{sec:intro}
Sub-PeV Galactic diffuse gamma-ray emission (sub-PeV GDE, $E>100\,{\rm TeV}$) are first detected by the Tibet AS$\gamma$ experiment, later followed by the LHAASO observatory \citep{TibetDiffuse, PhysRevLett.131.151001}. However, the fluxes of sub-PeV GDE measured by these two experiments (hereafter called the {\it Tibet diffuse flux} and the {\it LHAASO diffuse flux}) significantly differ; the Tibet diffuse flux measured along the Galactic plane in the sky region of $25^{\circ}<l<100^{\circ}$ and $|b|<5^{\circ}$ is about five times higher than the LHAASO diffuse flux in the sky region of $15^{\circ}<l<125^{\circ}$ and $|b|<5^{\circ}$ when they are evaluated in the 100 TeV energy range. There are two possible interpretations to explain the origin of the difference. The first interpretation is that LHAASO masks a significant portion of the sky to remove the contamination from resolved gamma-ray sources, which could lead to the underestimation of diffuse gamma-ray flux, as mentioned by \cite{Fang_2023_ApJL}. This interpretation is supported by the fact that the Tibet diffuse flux is consistent with the theoretical prediction given by \cite{Lipari_and_Vernetto_PRD_2018}. The second interpretation is that the Tibet diffuse flux is largely contaminated by emissions from Galactic gamma-ray sources because of its insufficient source-masking scheme, as pointed out by \cite{Yan_et_al_NatAstron_2024}. Some theoretical studies also suggest that a large fraction of the Tibet diffuse flux could be the contribution from gamma-ray sources, including unresolved ones \citep{Liu_2021, Vecchiotti_2022}. However, if one considers only resolved Galactic gamma-ray sources, the contribution of such sources to the Tibet diffuse flux can be quantified using the first LHAASO catalog (1LHAASO catalog, \citealp{1LHAASOCatalog}), the most complete gamma-ray source catalog to date in the northern sky covering the sub-PeV energy range. In fact, \cite{Kato_2024} demonstrate that none of the $23$ Tibet diffuse gamma-ray events detected above $398\, {\rm TeV}$ come from the 1LHAASO catalog sources detected above 100 TeV, supporting the diffusive nature of these events.

\cite{Fang_2023_ApJL} compare the Tibet diffuse flux with the total gamma-ray flux of the 1LHAASO catalog sources detected above $25\,{\rm TeV}$; see Figure 2 of their paper. Their comparison shows that the source flux significantly contributes to the Tibet diffuse flux ($\sim 50\%$ in the 100 TeV energy range) in the sky region of $25^{\circ}<l<100^{\circ}$ and $|b|<5^{\circ}$, while the Tibet diffuse flux may dominate the source flux in the sky region of $50^{\circ}<l<200^{\circ}$ and $|b|<5^{\circ}$. However, their study should be revised from the following three aspects: first, the fraction of the source flux in the Tibet diffuse flux should be quantitatively concluded; second, the fact that some of the 1LHAASO catalog sources are partially masked in the measurement of the Tibet diffuse flux should be properly taken into account; third, the contribution of the Cygnus Cocoon \citep{HAWC_Cygnus, LHAASOCOLLABORATION2024449} should be included into the source flux. Accounting for the latter two aspects, one should quantify the contribution of the source flux to the Tibet diffuse flux. This would give us rich implications about the difference between the Tibet and LHAASO diffuse fluxes.

\section{Results} \label{sec:results}
\begin{figure}
  \centering
  \includegraphics[scale=1.0]{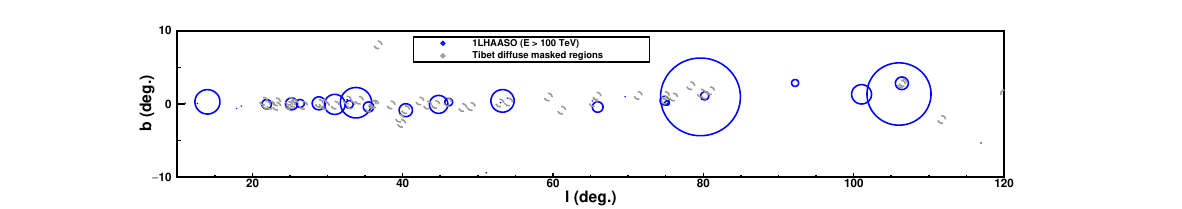}
  \includegraphics[scale=1.0]{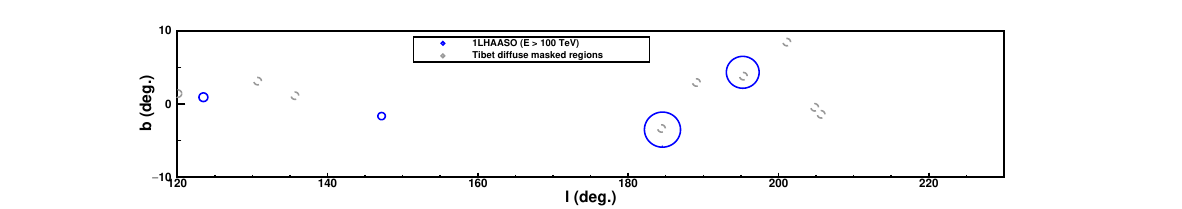}
  \caption{Positions and extensions of the sub-PeV LHAASO sources and the Cygnus Cocoon (blue circles) shown in the Galactic coordinates. The extensions have the $95\%$ containment radii. The point-like sources with only upper limits on the extensions are plotted with the blue dots. The large circle with a radius of $\simeq 5^{\circ}$ centered at $(l,b)\simeq (79{\fdg}6, 1{\fdg}2)$ shows the extension of LHAASO J2027+4119 to represent that of the Cygnus Cocoon; see the text. The gray circles with a fixed radius of $0{\fdg}5$ enclose the sky regions masked in the Tibet diffuse analysis \citep{TibetDiffuse}.}
  \label{LHAASOsources_and_mask}
\end{figure}

Our study focuses on the 1LHAASO catalog sources detected above $100\, {\rm TeV}$ with test statistics higher than $20$ (hereafter called the {\it sub-PeV LHAASO sources}) plus the Cygnus Cocoon. Figure \ref{LHAASOsources_and_mask} shows the positions and extensions of the sub-PeV LHAASO sources and the Cygnus Cocoon with the blue circles or dots. The circles have extensions with $95\%$ containment radii, which are calculated from the $39\%$ containment radii published by \cite{1LHAASOCatalog} and \cite{LHAASOCOLLABORATION2024449} assuming simple Gaussian source morphologies. The point-like sources with only upper limits on the extensions are plotted with the dots. For the Cygnus Cocoon, the extension with a $95\%$ containment radius of LHAASO J2027$+$4119 is plotted; see \cite{LHAASOCOLLABORATION2024449}. The gray circles with a fixed radius of $0{\fdg}5$ enclose the sky regions masked in the diffuse gamma-ray analysis by Tibet AS$\gamma$ (Tibet diffuse analysis, \citealp{TibetDiffuse}), and the masked source are shown in Table \ref{tab1}. The figure shows that many of the gamma-ray sources are totally or partially situated within the masked regions, implying that a significant fraction of the gamma-ray fluxes of these sources are masked in the Tibet diffuse analysis. This is not surprising because the Tibet diffuse analysis masked the gamma-ray sources registered in the TeVCat catalog\footnote{http://tevcat.uchicago.edu.} as of 2021, and many of the sub-PeV LHAASO sources are spatially associated with these TeVCat catalog sources.

The gamma-ray fluxes of the sub-PeV LHAASO sources are calculated by making use of the results given by \cite{1LHAASOCatalog}. \cite{1LHAASOCatalog} assume a simple power-law function for the gamma-ray energy spectra of the sub-PeV LHAASO sources and give a flux $N_0$ normalized at $50\, {\rm TeV}$ and differential spectral index $\Gamma$ for each of the sources from the maximum likelihood analysis applied to the data recorded with the LHAASO KM2A above 25 TeV. In order to model the morphology and spectrum of the gamma-ray emission of the Cygnus Cocoon, \cite{LHAASOCOLLABORATION2024449} simultaneously fits the following two templates to the data recorded with the LHAASO KM2A above 25 TeV: a Gaussian template whose gamma-ray emission is named as LHAASO J2027$+$4119, and a molecular-cloud template. \cite{LHAASOCOLLABORATION2024449} assumes a simple power-law function for the gamma-ray energy spectra of the two templates and gives the best-fit parameters from the maximum likelihood analysis. Our study estimates the gamma-ray flux of the Cygnus Cocoon by summing up the fluxes calculated from the best-fit power-law functions to the two templates.

To constrain the fraction of the gamma-ray fluxes of the sub-PeV LHAASO sources that are not masked in the Tibet diffuse analysis, a simple toy Monte Carlo (MC) simulation is performed. A total $10^5$ random events are generated following a Gaussian with a width of $r = \sqrt{r^{2}_{39\%}+r^{2}_{\rm tibPSF}}$ with respect to the centroid of a source of interest. Here $r_{39\%}$ is the extension of the source and $r_{39\%} = 0^{\circ}$ for point-like sources which have only upper limits on the extensions. $r_{\rm tibPSF}$ is the $39\%$ containment radius of the point spread function of the Tibet Air Shower Array for sub-PeV gamma rays and is fixed at $0{\fdg}2$; see the supplementary material of \cite{TibetDiffuse}. The number $N_{\rm masked}$ of the random events generated inside the sky regions masked in the Tibet diffuse analysis is counted, and the following coefficient $\alpha$ is calculated: $\alpha = (N_{\rm tot} - N_{\rm masked}) / N_{\rm tot}$, where $N_{\rm tot} = 10^5$ is the total number of the generated random events. The toy MC simulations are performed for all the sub-PeV LHAASO sources and Cygnus Cocoon and the corresponding $\alpha$'s are calculated.

The total gamma-ray flux $F_{\rm src,\, tot}$ of the resolved sources contributing to the Tibet diffuse flux is estimated as
\begin{equation}
  F_{\rm src,\, tot}(E) = \sum_{i} \alpha_{i} N_{0,\,i} \bigg(\frac{E}{50\,{\rm TeV}}\bigg)^{-\Gamma_{i}}
\end{equation}
where the summation runs over the sub-PeV LHAASO sources and the Cygnus Cocoon. Hereafter, $F_{\rm src,\, tot}(E)$ is called the source flux. The error of the source flux is calculated following the methodology used by \cite{Fang_2023_ApJL}. First, the statistical error of the source flux is calculated by summing up the statistical errors of $N_{0}$'s in quadrature for error propagation. Then, the systematic error of $7\%$ of the flux \citep{1LHAASOCatalog} is added to the statistical error in quadrature, leading to the total error of the source flux. Note that the above calculation should give us an upper limit on the source flux in the sub-PeV energy range. The $N_{0}$'s and $\Gamma$'s are given from the analysis of the data recorded with the LHAASO KM2A above 25 TeV and their values should mainly reflect the characteristics of the gamma-ray spectra in the several tens of TeV energy range. On the other hand, the sub-PeV LHAASO sources also detected by the LHAASO WCD in $1\,{\rm TeV}<E<25\,{\rm TeV}$ show a significant steepening between the gamma-ray energy spectra measured with the WCD and KM2A. This fact makes us expect that the spectrum is further steeper in the sub-PeV energy range and the true sub-PeV flux is lower than our calculation based on the results of the measurement above 25TeV.

Figure \ref{comparison} compares the source flux shown with the blue curve with the Tibet diffuse flux shown with the orange points in the two sky regions: region A ($25^{\circ}<l<100^{\circ}$ and $|b|<5^{\circ}$) and region B ($50^{\circ}<l<200^{\circ}$ and $|b|<5^{\circ}$) studied in the Tibet diffuse analysis \citep{TibetDiffuse}. The terminology of these regions follows \cite{Fang_2023_ApJL}. The blue shaded band shows the total error of the source flux. The statistical error of the Tibet diffuse flux is shown with the orange vertical bars, while the yellow vertical bars show the total error of the statistical and systematic (30\%, \citealp{TibetDiffuse}) errors summed up in quadrature. Table \ref{tab2} shows the fraction of the source flux in the Tibet diffuse flux evaluated at the representative energies of the three energy bins adopted in the Tibet diffuse analysis. The error of the fraction is calculated from the total errors, including the systematic errors, of the source flux and the Tibet diffuse flux for error propagation. The result shows that in region A, the source flux accounts for less than $26.9\%\pm 9.9\%$, $34.8\%\pm 14.0\%$, and $13.5\%^{+6.3\%}_{-7.7\%}$ of the Tibet diffuse flux at $121\, {\rm TeV}$, $220\, {\rm TeV}$, and $534\, {\rm TeV}$, respectively. In region B, the fraction is less than $24.1\%\pm 9.5\%$, $27.1\%\pm 11.1\%$, and $13.5\%^{+6.2\%}_{-7.6\%}$. These upper limits are consistent with the estimate made by \cite{TibetDiffuse} for the fractional source contribution of $13\%$ to the Tibet diffuse flux above 100 TeV. In particular, the upper limits at $534\, {\rm TeV}$ are consistent with the picture presented by \cite{Kato_2024} which support the diffusive nature of the diffuse gamma-ray events observed by Tibet AS$\gamma$ above $398\, {\rm TeV}$. The resultant upper limits for regions A and B are plus $8\%$ at maximum when all the 1LHAASO catalog sources detected by the LHAASO KM2A above 25 TeV, instead of above 100 TeV, are included in the calculation of the source flux (see Appendix \ref{app1}).

The gray lines in Figure \ref{comparison} show the best-fit energy spectra of some sources which largely contribute to the source flux in the sub-PeV energy range. The Cygnus Cocoon gives the largest contribution in both regions A and B; it accounts for $34.9\%$ ($43.6\%$) of the source flux at 100 TeV (1 PeV) in region A, while $35.8\%$ ($51.8\%$) at 100 TeV (1 PeV) in region B. Half the source flux is dominated by the Cygnus Cocoon and 1LHAAO J1908$+$0615u at 100 TeV to 1 PeV in region A. In region B, half the source flux is dominated by the Cygnus Cocoon and 1LHAASO J2229$+$5927u at 100 TeV, and by the Cygnus Cocoon and 1LHAASO J2228$+$6100u at 1 PeV. The contributions of the sources to the source flux are summarized in Tables \ref{tab_relative_contributionA} and \ref{tab_relative_contributionB} in Appendix \ref{app2}.

It is interesting to estimate the fraction of the source flux plus the LHAASO diffuse flux in the Tibet diffuse flux. LHAASO measured the energy spectra of GDE in the inner ($15^{\circ}<l<125^{\circ}$ and $|b|<5^{\circ}$) and outer ($125^{\circ}<l<235^{\circ}$ and $|b|<5^{\circ}$) Galactic plane regions, which are different from regions A and B studied in the Tibet diffuse analysis. Nonetheless, it is not inappropriate to compare the LHAASO diffuse flux in the inner Galactic plane region and the Tibet diffuse flux in region A because the inner Galactic plane region includes region A, and their longitudinal ranges do not differ much. Using the best-fit power-law function to the energy spectrum of GDE observed by LHAASO in the inner Galactic plane region \citep{PhysRevLett.131.151001}, the ratio of the LHAASO diffuse flux to the Tibet diffuse flux is evaluated as $22.5\%$, $30.7\%$, and $12.3\%$ at $121\, {\rm TeV}$, $220\, {\rm TeV}$, and $534\, {\rm TeV}$, respectively. As a result, the source flux in region A plus the LHAASO diffuse flux in the inner Galactic plane region accounts for less than $49.4\%$, $65.5\%$, and $25.8\%$ of the Tibet diffuse flux in region A at $121\, {\rm TeV}$, $220\, {\rm TeV}$, and $534\, {\rm TeV}$, respectively. This means that at least in region A, the total gamma-ray flux of the currently resolved sub-PeV gamma-ray sources, i.e., the sub-PeV LHAASO sources plus the Cygnus Cocoon, cannot completely account for the difference between the Tibet and LHAASO diffuse fluxes in the sub-PeV energy range.

\section{Discussions} \label{sec:dis}
The Tibet diffuse flux is well consistent with the Galactic neutrino flux measured by the IceCube Collaboration \citep{IceCube_2023} based on the $\pi^0$ template \citep{Ackermann_2012}, supporting that the Tibet diffuse flux is dominated by the gamma-ray emission of hadronic origin. This perspective is also supported by \cite{Fang_2021}; the authors predicted the Galactic diffuse neutrino flux from the Tibet diffuse flux, which is found to be consistent with the IceCube observation. The source flux of the sub-PeV LHAASO sources associated with pulsars, which are likely the gamma-ray sources of leptonic origin (see Table 4 of \citealp{1LHAASOCatalog}), only accounts for less than $8.9\%$, $11.8\%$, and $4.7\%$ of the Tibet diffuse flux in region A at $121\,{\rm TeV}$, $220\, {\rm TeV}$, and $534\, {\rm TeV}$, respectively. In region B, the fraction is less than $6.6\%$, $6.9\%$, and $3.0\%$. Furthermore, the contribution of unresolved leptonic sources can be constrained from the data-driven study performed by \cite{Samy_arXiv_v1_2024}. The authors estimate the total source flux of unresolved leptonic sources potentially associated with pulsars and find that the flux reaches $5\times 10^{-16}\,{\rm TeV}^{-1}\,{\rm cm}^{-2}\, {\rm s}^{-1}\, {\rm sr}^{-1}$ at 120 TeV in the inner Galactic plane region observed by LHAASO when no masking scheme is adopted. Therefore, their estimate can put a stringent upper limit on the contribution of unresolved leptonic sources to the Tibet diffuse flux in region A. Figure \ref{comparison_unres_lep} compares the Tibet diffuse flux with the total source flux of the unresolved leptonic sources presented by \cite{Samy_arXiv_v1_2024}, and the latter only contributes less than $16.7\%$, $22.1\%$, and $8.1\%$ of the former at $121\,{\rm TeV}$, $220\, {\rm TeV}$, and $534\, {\rm TeV}$, respectively.

Moreover, \cite{Fang_2023_ApJL} claim that the Galactic neutrino flux is likely dominated by the diffuse emission and/or the emission from unresolved hadronic sources. Therefore, if the Galactic neutrino flux is dominated by the diffuse neutrino emission, the aforementioned studies \citep{IceCube_2023, Fang_2023_ApJL, Samy_arXiv_v1_2024} can conclude that the Tibet diffuse flux is dominated by diffuse gamma-ray emission of hadronic origin. Note that this perspective is supported by the fact that the Tibet diffuse flux, even after subtracting the source flux constrained in our study, is consistent with the theoretical model of the sub-PeV GDE flux given by \cite{Lipari_and_Vernetto_PRD_2018} both in regions A and B. The consistency gives us a natural interpretation of diffusive nature of the Tibet diffuse flux.

On the other hand, the hadronic nature of GDE observed by LHAASO, especially above 30 TeV, is indicated by \cite{Fang_2023_ApJL}. Their result is also supported by \cite{Samy_arXiv_v1_2024}, in which the authors estimate the contribution to the LHAASO diffuse flux from the leptonic gamma-ray emission of unresolved sources associated with pulsars as less than 20\% in the sub-PeV energy range.

After subtracting the source flux, the Tibet diffuse flux in region A is higher than the LHAASO diffuse flux in the inner Galactic plane region by about three, two, and seven times at $121\, {\rm TeV}$, $220\, {\rm TeV}$, and $534\, {\rm TeV}$, respectively. Indeed, the difference could be explained by the scenario that Tibet AS$\gamma$ and LHAASO observe sub-PeV GDE of hadronic origin in different Galactic latitudinal regions due to their different source-masking schemes. To validate the scenario, the theoretical model given by \cite{Lipari_and_Vernetto_PRD_2018} is used. The authors present the distribution of GDE at 12 GeV in the Galactic latitude which well reproduces the Fermi-LAT data \citep{Ackermann_2012}; see Figure 8 of their paper. Since LHAASO masks a significant portion of the sky region in $|b|<2^{\circ}$ in the inner Galactic plane region (see the upper panel of Figure 1 of \citealp{PhysRevLett.131.151001}), one can approximately consider that they effectively measure the diffuse gamma-ray flux in $2^{\circ}<|b|<5^{\circ}$. On the other hand, the Tibet AS$\gamma$ only masks less than $10\%$ of region A, and one can approximately consider that they appropriately measure the full diffuse flux in $|b|<5^{\circ}$. The calculation based on the latitude distribution of GDE of \cite{Lipari_and_Vernetto_PRD_2018} shows that the diffuse gamma-ray flux per solid angle in region A is three times higher than that in the inner Galactic plane region with $2^{\circ}<|b|<5^{\circ}$; the result is consistent with the difference between the Tibet and LHAASO diffuse fluxes.


\section{Conclusion} \label{sec:con}
Our study constrains the contribution of the source flux, the gamma-ray flux of the sub-PeV gamma-ray sources currently resolved by LHAASO (those registered in the first LHAASO catalog \citep{1LHAASOCatalog} plus the Cygnus Cocoon \citep{LHAASOCOLLABORATION2024449}), to the Tibet diffuse flux, taking into account the source-masking scheme adopted in the Tibet diffuse analysis. The fraction of the source flux in the Tibet diffuse flux is found to be subdominant; it is less than $26.9\%\pm 9.9\%$, $34.8\%\pm 14.0\%$, and $13.5\%^{+6.3\%}_{-7.7\%}$ in region A ($25^{\circ}<l<100^{\circ}$ and $|b|<5^{\circ}$) at $121\, {\rm TeV}$, $220\, {\rm TeV}$, and $534\, {\rm TeV}$, respectively. Therefore, the currently resolved sub-PeV gamma-ray sources cannot completely account for the five times difference at 100 TeV between the Tibet diffuse flux in region A and LHAASO diffuse flux in the inner Galactic Plane region ($15^{\circ}<l<125^{\circ}$ and $|b|<5^{\circ}$). In region B ($50^{\circ}<l<200^{\circ}$ and $|b|<5^{\circ}$), the fraction is less than $24.1\%\pm 9.5\%$, $27.1\%\pm 11.1\%$, and $13.5\%^{+6.2\%}_{-7.6\%}$. From the previous IceCube neutrino observation \citep{IceCube_2023}, data-driven studies \citep{Fang_2021, Fang_2023_ApJL, Samy_arXiv_v1_2024}, and a theoretical GDE model \citep{Lipari_and_Vernetto_PRD_2018}, the hadronic diffusive nature of the Tibet diffuse flux is the most natural interpretation after subtracting the source flux, although some contributions from very faint unresolved hadronic gamma-ray sources cannot be ruled out. Note that \cite{Fang_2023_ApJL} and \cite{Samy_arXiv_v1_2024} also support the hadronic nature of the sub-PeV LHAASO diffuse flux. Different source-masking schemes adopted by Tibet AS$\gamma$ and LHAASO for their diffuse analyses result in different effective galactic latitudinal ranges of the sky regions observed by the two experiments. Our study concludes that the effect of the different source-masking schemes leads to the observed difference (about three times in the sub-PeV energy range) between the Tibet diffuse flux measured in region A and LHAASO diffuse flux in the inner Galactic plane region, even though both experiments observe sub-PeV GDE of hadronic origin.

\begin{acknowledgments}
  We thank an anonymous reviewer for her or his careful reading and helpful comments that contributed to improve our article. This work is supported in part by Grants-in-Aid for Scientific Research from the Japan Society for the Promotion of Science in Japan, the joint research program of the Institute for Cosmic Ray Research (ICRR), the University of Tokyo, and the use of the computer system of ICRR. This work is also supported by the National Natural Science Foundation of China under Grants No. 12227804, and the Key Laboratory of Particle Astrophysics, Institute of High Energy Physics, CAS.
\end{acknowledgments}

\clearpage
\startlongtable
\begin{deluxetable*}{crrr}
\tablenum{1}
\tablecaption{Sky regions masked in the Tibet diffuse analysis \label{tab1}}
\tablehead{\colhead{$l$ ($^{\circ}$)} & \colhead{$b$ ($^{\circ}$)} & \colhead{Masked source} & \colhead{Reference}}
\startdata
120.10 &1.41 &Tycho & \cite{Archambault_2017}\\
130.71 &3.11 &3C 58 & \cite{MAGIC_2014}\\
135.67 &1.12 &LS I +61 303 & \cite{doi:10.1126/science.1128177}\\
184.44 &-3.36 &HAWC J0543+233 & The Astronomer's Telegram\footnote[1]{https://www.astronomerstelegram.org/?read=10941}\\
189.08 &2.93 &IC 443 & \cite{Acciari_2009}\\
195.34 &3.79 &Geminga & \cite{Abdo_2009}\\
205.68 &-1.43 &HESS J0632+057 & \cite{HGPS2018}\\
204.82 &-0.47 &HAWC J0635+070 & The Astronomer's Telegram\footnote[2]{https://www.astronomerstelegram.org/?read=12013}\\
201.11 &8.45 &2HWC J0700+143 & \cite{2HWCCatalog}\\
21.50 &0.37 &HESS J1828-099 & \cite{HGPS2018}\\
36.72 &8.08 &2HWC J1829+070 & \cite{2HWCCatalog}\\
21.86 &-0.12 &HESS J1831-098 & \cite{HGPS2018}\\
23.21 &0.29 &HESS J1832-085 & \cite{HGPS2018}\\
22.48 &-0.19 &HESS J1832-093 & \cite{HGPS2018}\\
23.25 &-0.32 &HESS J1834-087 & \cite{HGPS2018}\\
24.94 &0.36 &MAGIC J1835-069 & \cite{10.1093/mnras/sty3387}\\
24.86 &-0.21 &MAGIC J1837-073 & \cite{10.1093/mnras/sty3387}\\
25.49 &0.09 &2HWC J1837-065 & \cite{2HWCCatalog}\\
25.18 &-0.12 &HESS J1837-069 & \cite{HGPS2018}\\
26.80 &-0.21 &HESS J1841-055 & \cite{HGPS2018}\\
28.90 &0.07 &HESS J1843-033 & \cite{HGPS2018}\\
29.42 &0.08 &HESS J1844-030 & \cite{HGPS2018}\\
29.72 &-0.24 &HESS J1846-029 & \cite{HGPS2018}\\
31.01 &-0.17 &HESS J1848-018 & \cite{HGPS2018}\\
32.62 &0.53 &IGR J18490-0000 & \cite{HGPS2018}\\
34.23 &0.49 &2HWC J1852+013 & \cite{2HWCCatalog}\\
33.12 &-0.14 &HESS J1852-000 & \cite{HGPS2018}\\
35.97 &-0.07 &HESS J1857+026 & \cite{HGPS2018}\\
36.28 &-0.02 &MAGIC J1857.6+0297 & \cite{MAGIC_J1857.6+0297}\\
35.58 &-0.59 &HESS J1858+020 & \cite{HGPS2018}\\
38.47 &-0.14 &2HWC J1902+048 & \cite{2HWCCatalog}\\
42.29 &0.40 &2HWC J1907+084 & \cite{2HWCCatalog}\\
40.39 &-0.79 &MGRO J1908+06 & \cite{HESS_MGRO1908}\\
39.61 &-1.96 &SS 433 w1 & \cite{SS_443}\\
43.32 &-0.17 &W 49B & \cite{HESS_W49B}\\
44.40 &-0.08 &HESS J1912+101 & \cite{HGPS2018}\\
39.86 &-2.67 &SS 433 e1 & \cite{SS_443}\\
46.00 &0.24 &2HWC J1914+117 & \cite{2HWCCatalog}\\
47.99 &-0.50 &2HWC J1921+131 & \cite{2HWCCatalog}\\
49.12 &-0.37 &W 51 & \cite{MAGIC_W51}\\
52.93 &0.13 &2HWC J1928+177 & \cite{2HWCCatalog}\\
54.10 &0.25 &SNR G054.1+00.3 & \cite{Acciari_2010}\\
59.38 &0.94 &2HWC J1938+238 & \cite{2HWCCatalog}\\
61.16 &-0.86 &2HWC J1949+244 & \cite{2HWCCatalog}\\
65.86 &1.06 &2HWC J1953+294 & \cite{2HWCCatalog}\\
65.35 &0.18 &2HWC J1955+285 & \cite{2HWCCatalog}\\
71.33 &1.16 &2HWC J2006+341 & \cite{2HWCCatalog}\\
74.95 &1.14 &VER J2016+371 & \cite{Aliu_2014}\\
74.93 &0.51 &MGRO J2019+37 & \cite{Abdo_2012}\\
76.05 &0.94 &Milagro Diffuse & \cite{Abdo_2008}\\
78.33 &2.48 &VER J2019+407 & \cite{Aliu_2013}\\
79.72 &1.47 &MGRO J2031+41 & \cite{Abdo_2012}\\
80.96 &1.80 &ARGO J2031+4157 & \cite{Bartoli_2014}\\
80.25 &1.20 &TeV J2032+4130 & \cite{Abeysekara_2018}\\
106.35 &2.71 &SNR G106.3+02.7 & \cite{Acciari_2009_2}\\
106.58 &2.91 &Boomerang & \cite{Abdo_2009}\\
111.72 &-2.13 &Cassiopeia A & \cite{MAGIC_CassiopeiaA}\\
\enddata
\tablecomments{Each of the masked regions has a fixed radius of $0{\fdg}5$. The first and second columns show the center of the masked region in the Galactic longitude and latitude, respectively. The third column shows the masked gamma-ray source which is registered in the TeVCat catalog. The fourth column gives the reference which determines the coordinates of the masked source position presented in the TeVCat catalog.}
\end{deluxetable*}
  

\begin{figure}
  \centering
  \includegraphics[scale=0.4]{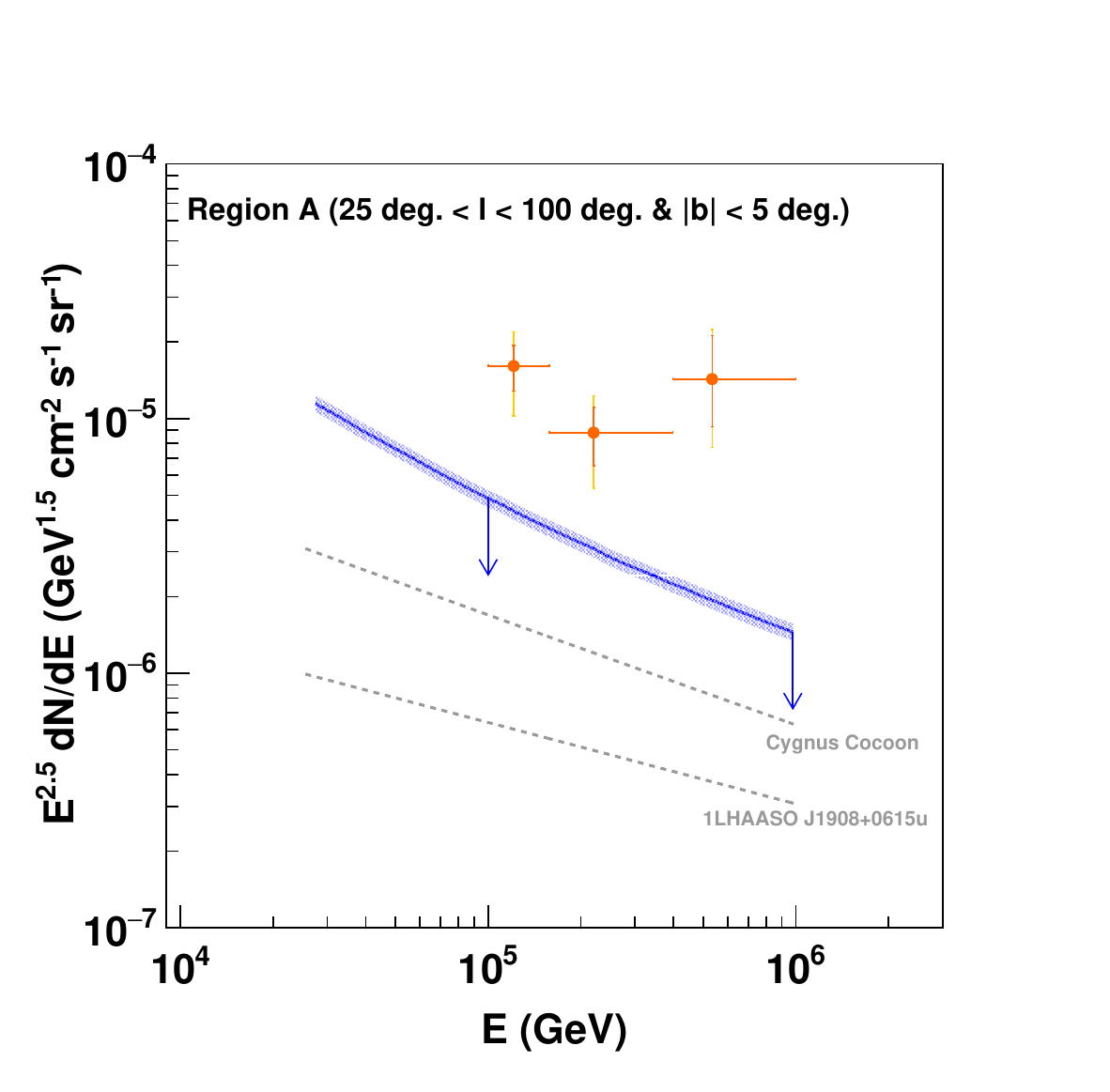}
  \includegraphics[scale=0.4]{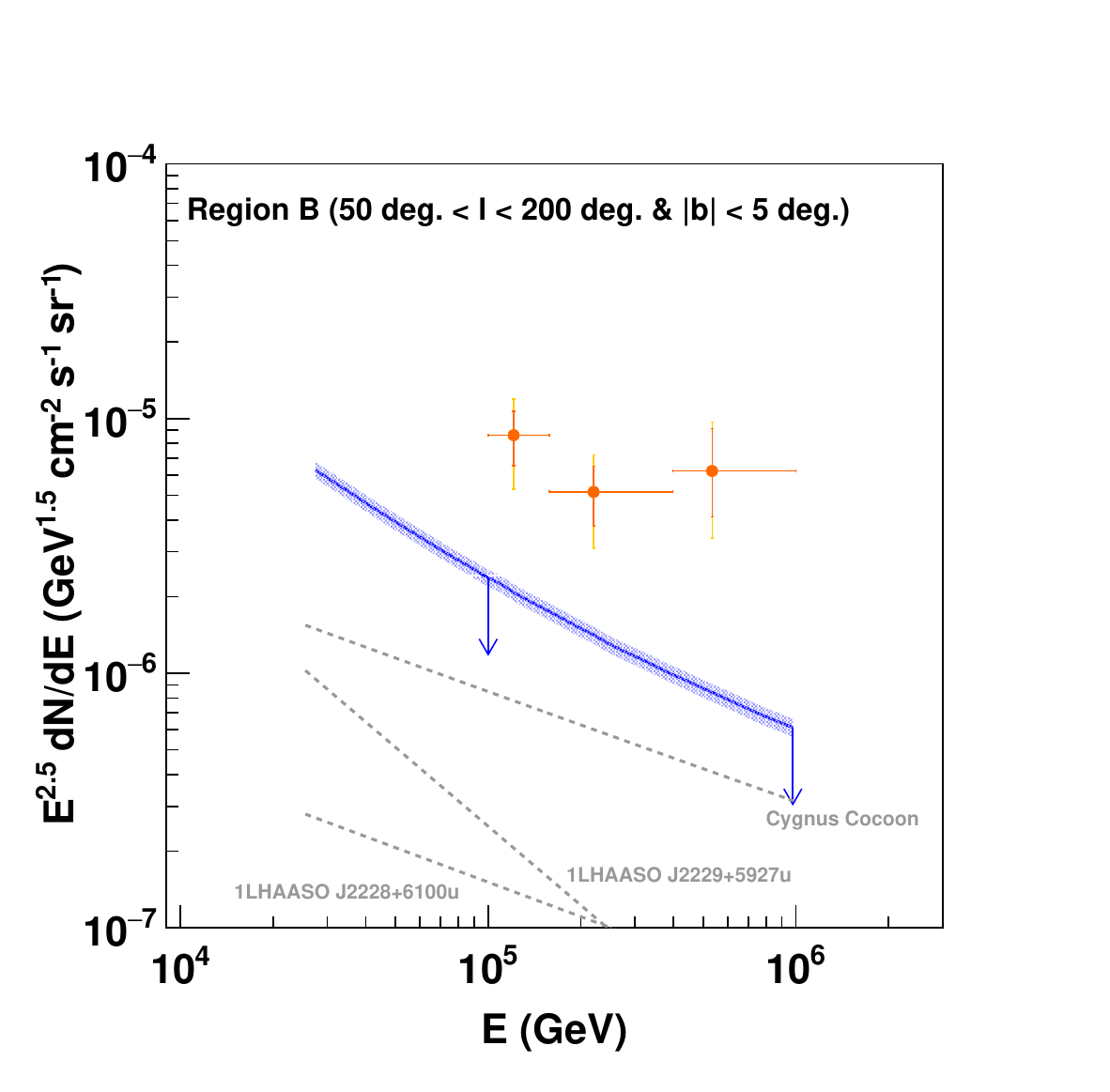}
  \caption{Tibet diffuse flux (orange points) compared with the total gamma-ray flux of the sub-PeV LHAASO sources plus the Cygnus Cocoon that is not masked in the Tibet diffuse analysis (source flux, blue curve). The comparison is made for regions A (left panel) and B (right panel). The statistical error of the Tibet diffuse flux is shown with the orange vertical bars, while the yellow vertical bars show the quadrature sum of the statistical and systematic errors. The blue shaded band shows the error of the source flux; see the text for detailed descriptions of the error calculation. The downward arrows at 100 TeV and 1 PeV indicate that the calculated source flux should be regarded as the upper limit in the sub-PeV energy range; see the text. The gray lines show the best-fit energy spectra of some sources which largely contribute to the source flux in the sub-PeV energy range; see the text and Appendix \ref{app2}.}
  \label{comparison}

\end{figure}
\begin{figure}
  \centering
  \includegraphics[scale=0.4]{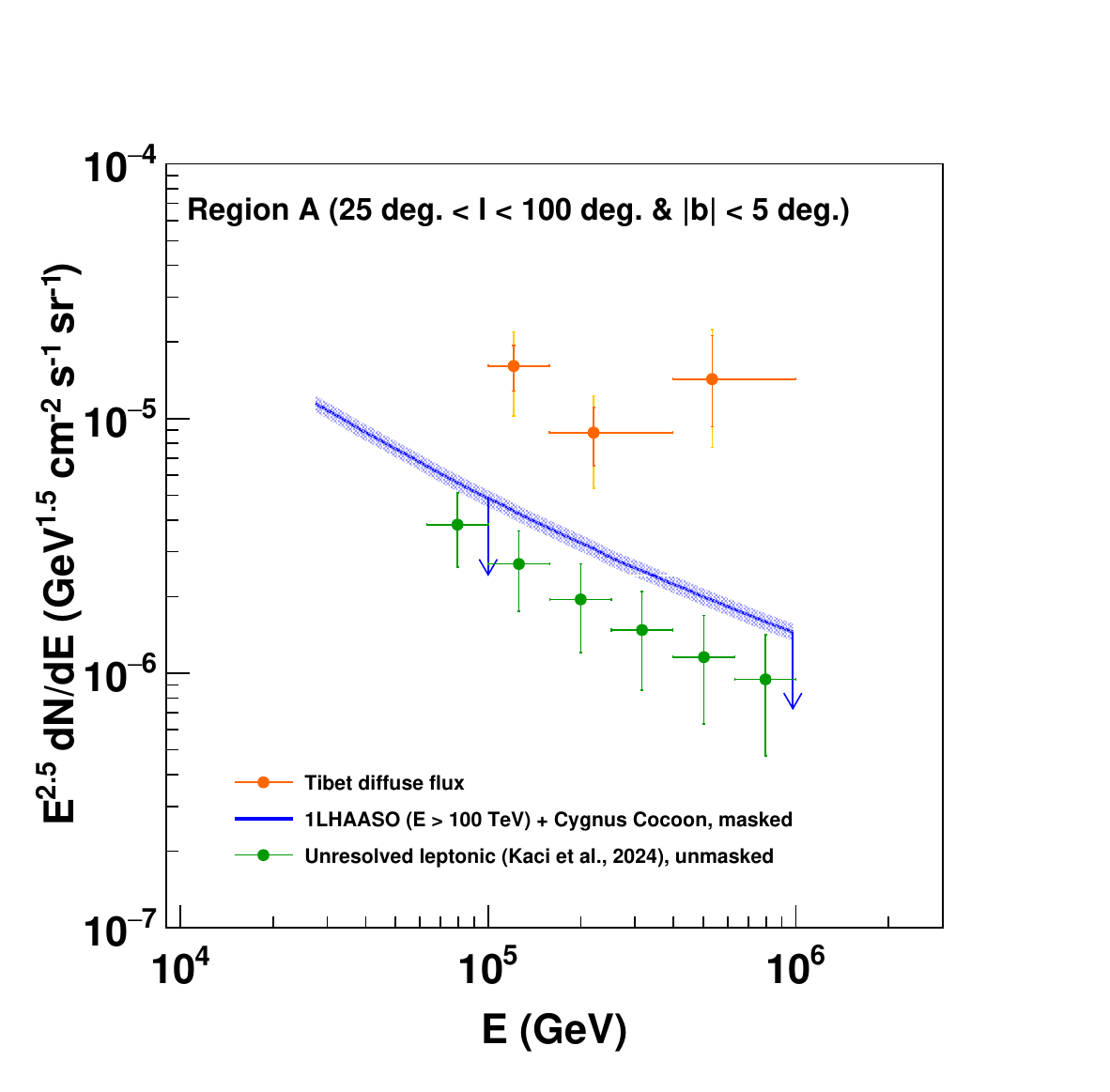}
  \caption{Same as the left panel of Figure \ref{comparison}, but additionally shows the source flux of unresolved leptonic sources estimated by \cite{Samy_arXiv_v1_2024} (green points). This leptonic source flux is estimated for the inner Galactic plane region observed by LHAASO when no masking scheme is adopted. Therefore, the flux should be regarded as a stringent upper limit on the contribution of unresolved leptonic sources to the Tibet diffuse flux in region A. The vertical error bars of the green point indicate the standard deviation obtained for the 100 simulations performed by \cite{Samy_arXiv_v1_2024}.}
  \label{comparison_unres_lep}
\end{figure}

\clearpage
\begin{deluxetable*}{lcccc}
\tablenum{2}
\tablecaption{Fraction of the source flux in the Tibet diffuse flux in regions A and B\label{tab2}}
\tablehead{\colhead{Region} & \colhead{$E$} & \colhead{Tibet diffuse flux} & \colhead{Source flux} & \colhead{Source flux/Tibet diffuse flux} \\
           \colhead{} & \colhead{(TeV)} & \colhead{(${\rm TeV}^{-1}\, {\rm cm}^{-2}\, {\rm s}^{-1}\, {\rm sr}^{-1}$)} & \colhead{(${\rm TeV}^{-1}\, {\rm cm}^{-2}\, {\rm s}^{-1}\, {\rm sr}^{-1}$)} & \colhead{(\%)}}
\startdata
& $121$ & $(3.16\pm 0.64_{\rm stat}\pm 0.95_{\rm syst})\times 10^{-15}$ & $<(8.50 \pm 0.64)\times 10^{-16}$ & $<26.9\pm 9.9$ \\
A & $220$ & $(3.88\pm 1.00_{\rm stat}\pm 1.16_{\rm syst})\times 10^{-16}$ & $<(1.35 \pm 0.10)\times 10^{-16}$ & $<34.8\pm 14.0$ \\
& $534$ & $(6.86^{{+ 3.30}_{\rm stat}}_{{- 2.40}_{\rm stat}}\pm 2.06_{\rm syst})\times 10^{-17}$ & $<(9.27 \pm 0.72)\times 10^{-18}$ & $<13.5^{+6.3}_{-7.7}$\\
\tableline
& $121$ & $(1.69\pm 0.41_{\rm stat}\pm 0.51_{\rm syst})\times 10^{-15}$ & $<(4.08 \pm 0.31)\times 10^{-16}$ & $<24.1\pm 9.5$ \\
B & $220$ & $(2.27\pm 0.60_{\rm stat}\pm 0.68_{\rm syst})\times 10^{-16}$ & $<(6.21 \pm 0.48)\times 10^{-17}$ & $<27.4\pm 11.1$ \\
& $534$ & $(2.99^{{+ 1.40}_{\rm stat}}_{{- 1.02}_{\rm stat}}\pm 0.90_{\rm syst})\times 10^{-17}$ & $<(4.03 \pm 0.32)\times 10^{-18}$ & $<13.5^{+6.2}_{-7.6}$ \\
\enddata
\tablecomments{The fraction is evaluated at the representative energies of the three energy bins, 121 TeV, 220 TeV, and 534 TeV, adopted in the Tibet diffuse analysis. The statistical and systematic errors are separately shown with subscripts {\it stat} and {\it syst} for the Tibet diffuse flux presented in the second column, while the error of the source flux presented in the third column includes both the statistical and systematic errors; see the text for details of the error calculation. The fourth column presents the fraction of the source flux in the Tibet diffuse flux. For the calculation of the error of the fraction, see the text}
\end{deluxetable*}


\clearpage
\appendix
\restartappendixnumbering
\section{Additional contribution from 1LHAASO catalog sources} \label{app1}
\begin{figure}[h]
  \centering
  \includegraphics[scale=0.4]{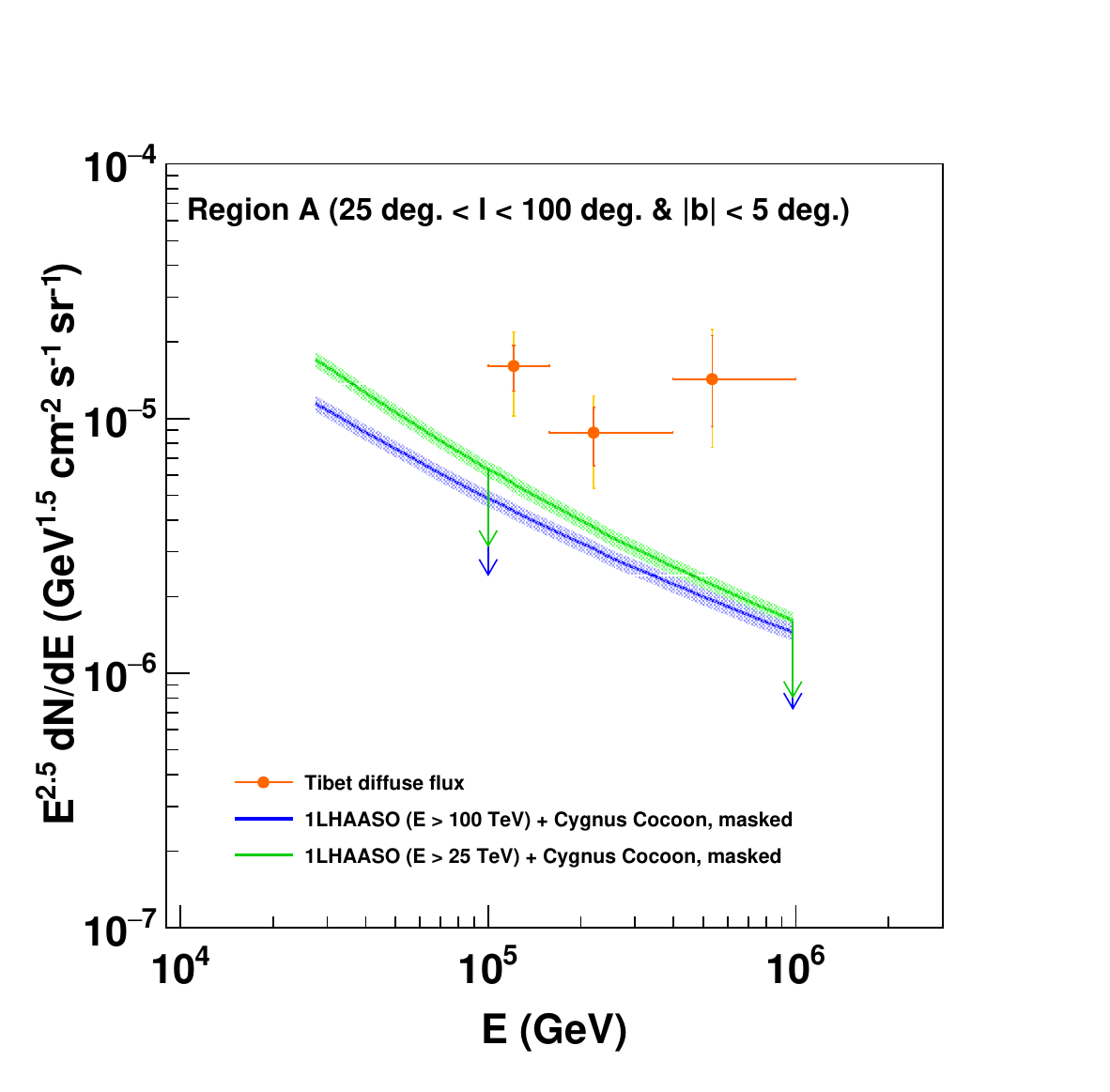}
  \includegraphics[scale=0.4]{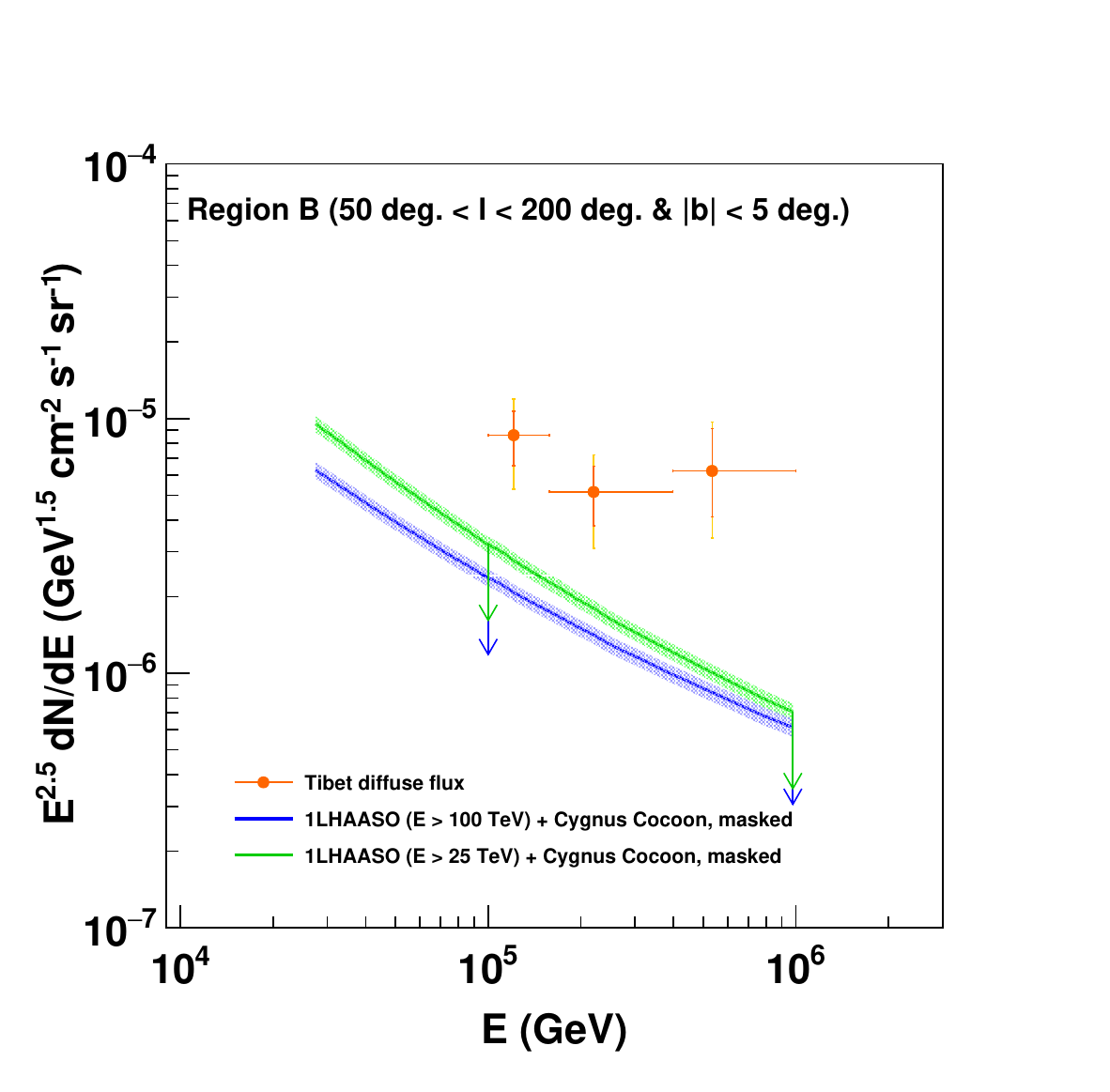}
  \caption{Same as Figure \ref{comparison}, but additionally shows the source flux including the contribution from all the 1LHAASO catalog sources detected above 25 TeV plus the Cygnus Cocoon (green curves and shaded bands).}
  \label{comparison_using_all_abv25TeV}
\end{figure}

Figure \ref{comparison_using_all_abv25TeV} is the same as Figure \ref{comparison}, but additionally shows, with the green curves, the source flux including the contribution from all the 1LHAASO catalog sources detected above 25 TeV plus the Cygnus Cocoon. The same method as presented in Section \ref{sec:results} is used to calculate the flux. Compared with the source flux only including the sub-PeV LHAASO sources and the Cygnus Cocoon, the flux increases approximately by $28\%$, $22\%$, and $15\%$ at $121\,{\rm TeV}$, $220\, {\rm TeV}$, and $534\, {\rm TeV}$, respectively, in region A. In region B, the flux increases approximately by $34\%$, $27\%$, $20\%$. These increases result in the fraction of the source flux in the Tibet diffuse flux of $34.5\%\pm 12.7\%$, $42.5\%\pm 17.1\%$, and $15.6\%^{+7.3\%}_{-8.9\%}$ in region A and $32.2\%\pm 12.7\%$, $34.8\%\pm 14.2\%$, and $16.1\%^{+7.4\%}_{-9.1\%}$ in region B at $121\,{\rm TeV}$, $220\, {\rm TeV}$, and $534\, {\rm TeV}$, respectively.

\section{Relative contributions of the sub-PeV LHAASO sources} \label{app2}
Tables \ref{tab_relative_contributionA} and \ref{tab_relative_contributionB} present the relative contributions of the sub-PeV LHAASO sources including the Cygnus Cocoon to the source flux in regions A and B. The relative contributions at 100 TeV and 1 PeV are evaluated.

\begin{deluxetable}{lrccc}
  \tablenum{3}
\tablecaption{Relative contributions of the sub-PeV LHAASO sources including the Cygnus Cocoon to the source flux at 100 TeV and 1 PeV in region A\label{tab_relative_contributionA}}
\tablehead{\colhead{Energy} & \colhead{Source name} & \colhead{Relative contribution} & \colhead{Cummulative contribution} \\
  \colhead{} & \colhead{} & \colhead{(\%)} & \colhead{(\%)}}
\startdata
&Cygnus Cocoon &34.9 &34.9\\
&1LHAASO J1908+0615u &13.2 &48.1 \\
&1LHAASO J1843$-$0335u &6.1 &54.2 \\
&1LHAASO J1839$-$0548u &5.2 &59.3 \\
&1LHAASO J1852+0050u &5.1 &64.4 \\
100 TeV &1LHAASO J1928+1813u &5.0 &69.4 \\
&1LHAASO J2108+5153u &5.0 &74.5 \\
&1LHAASO J1848$-$0153u &4.7 &79.2 \\
&1LHAASO J2020+3649u &3.5 &82.7 \\
&1LHAASO J1912+1014u &3.0 &85.7 \\
&1LHAASO J1959+2846u &2.9 &88.6 \\
&1LHAASO J2018+3643u &2.6 &91.2 \\
\tableline
&Cygnus Cocoon &43.6 &43.6 \\
&1LHAASO J1908+0615u &21.3 &64.9 \\
&1LHAASO J2108+5153u &5.8 &70.7 \\
&1LHAASO J1959+2846u &3.9 &74.6 \\
1 PeV &1LHAASO J2031+4052u &3.7 &78.2 \\
&1LHAASO J1839$-$0548u &3.2 &81.4 \\
&1LHAASO J1928+1813u &3.1 &84.5 \\
&1LHAASO J1843$-$0335u &2.4 &86.9 \\
&1LHAASO J2020+3649u &1.8 &88.7 \\
&1LHAASO J1912+1014u &1.8 &90.5 \\
\enddata
\tablecomments{The relative contributions of the sources are presented in descending order. {\it Cummulative contribution} is defined as the sum of the relative contributions of the sources with the largest contributions. The table presents the sources whose total flux accounts for more than 90\% of the source flux}
\end{deluxetable}

\begin{deluxetable}{lrccc}[h]
  \tablenum{4}
\tablecaption{Same as Table \ref{tab_relative_contributionA} but in region B\label{tab_relative_contributionB}}
\tablehead{\colhead{Energy} & \colhead{Source name} & \colhead{Relative contribution} & \colhead{Cummulative contribution} \\
  \colhead{} & \colhead{} & \colhead{(\%)} & \colhead{(\%)}}
\startdata
&Cygnus Cocoon &35.8 &35.8 \\
&1LHAASO J2229+5927u &10.6 &46.3 \\
&1LHAASO J0634+1741u &6.5 &52.9 \\
&1LHAASO J2228+6100u &6.4 &59.3 \\
&1LHAASO J0542+2311u &5.2 &64.5 \\
100 TeV &1LHAASO J1928+1813u &5.2 &69.7 \\
&1LHAASO J2108+5153u &5.2 &74.8 \\
&1LHAASO J2200+5643u &4.6 &79.4 \\
&1LHAASO J0056+6346u &4.3 &83.7 \\
&1LHAASO J2020+3649u &3.6 &87.3 \\
&1LHAASO J1959+2846u &3.0 &90.2 \\
\tableline
&Cygnus Cocoon &51.8 &51.8 \\
&1LHAASO J2228+6100u &8.9 &60.7 \\
&1LHAASO J2108+5153u &6.8 &67.5 \\
&1LHAASO J1959+2846u &4.6 &72.1 \\
1 PeV &1LHAASO J2031+4052u &4.4 &76.5 \\
&1LHAASO J2229+5927u &3.9 &80.4 \\
&1LHAASO J1928+1813u &3.7 &84.0 \\
&1LHAASO J0056+6346u &2.5 &86.5 \\
&1LHAASO J2020+3649u &2.2 &88.7 \\
&1LHAASO J2200+5643u &2.1 &90.8 \\
\enddata
\end{deluxetable}
  
\clearpage

\end{document}